\documentstyle[12pt]{article}

\begin{document}

\input{epsf}

\def\beq{\begin{equation}}
\def\eeq{\end{equation}}
\def\bea{\begin{eqnarray}}
\def\eea{\end{eqnarray}}
\def\beas{\begin{eqnarray*}}
\def\eeas{\end{eqnarray*}}
\def\ov{\overline}
\def\ot{\otimes}

\newcommand{\hf}{\mbox{$\frac{1}{2}$}}
\def\sig{\sigma}
\def\De{\Delta}
\def\af{\alpha}
\def\be{\beta}
\def\la{\lambda}
\def\ga{\gamma}
\def\ep{\epsilon}
\def\vep{\varepsilon}
\def\half{\frac{1}{2}}
\def\third{\frac{1}{3}}
\def\fth{\frac{1}{4}}
\def\sth{\frac{1}{6}}
\def\tth{\frac{1}{24}}
\def\tde{\frac{3}{2}}

\def\zb{{\bar z}} 
\def\psib{{\bar \psi}} 
\def\etab{{\bar \eta }}
\def\gab{{\bar \ga}}
\def\vev#1{\langle #1 \rangle}
\def\inv#1{{1 \over #1}}

\def\CA{{\cal A}}       \def\CB{{\cal B}}       \def\CC{{\cal C}}
\def\CD{{\cal D}}       \def\CE{{\cal E}}       \def\CF{{\cal F}}
\def\CG{{\cal G}}       \def\CH{{\cal H}}       \def\CI{{\cal J}}
\def\CJ{{\cal J}}       \def\CK{{\cal K}}       \def\CL{{\cal L}}
\def\CM{{\cal M}}       \def\CN{{\cal N}}       \def\CO{{\cal O}}
\def\CP{{\cal P}}       \def\CQ{{\cal Q}}       \def\CR{{\cal R}}
\def\CS{{\cal S}}       \def\CT{{\cal T}}       \def\CU{{\cal U}}
\def\CV{{\cal V}}       \def\CW{{\cal W}}       \def\CX{{\cal X}}
\def\CY{{\cal Y}}       \def\CZ{{\cal Z}}

\newcommand{\np}{Nucl. Phys.}
\newcommand{\pl}{Phys. Lett.}
\newcommand{\prl}{Phys. Rev. Lett.}
\newcommand{\cmp}{Commun. Math. Phys.}
\newcommand{\jmp}{J. Math. Phys.}
\newcommand{\jpamg}{J. Phys. {\bf A}: Math. Gen.}
\newcommand{\lmp}{Lett. Math. Phys.}
\newcommand{\ptp}{Prog. Theor. Phys.}

\newif\ifbbB\bbBfalse                
\bbBtrue                             

\ifbbB   
 \message{If you do not have msbm (blackboard bold) fonts,}
 \message{change the option at the top of the text file.}
 \font\blackboard=msbm10 
 \font\blackboards=msbm7 \font\blackboardss=msbm5
 \newfam\black \textfont\black=\blackboard
 \scriptfont\black=\blackboards \scriptscriptfont\black=\blackboardss
 \def\Bbb#1{{\fam\black\relax#1}}
\else
 \def\Bbb{\bf}
\fi

\def\bC{{\Bbb C}} 
\def\bZ{{\Bbb Z}}
\def\CN{{\cal N}}

\title{Non-unitary Conformal Field Theory and Logarithmic Operators for 
Disordered Systems}
\author{{\bf Z. Maassarani}\\
 \\
{\bf D. Serban}\\
 \\
{\small CEA-SACLAY, Service de Physique Th\'eorique}\\
{\small F-91191 Gif-sur-Yvette  Cedex, FRANCE}}
\date{}
\maketitle

\begin{abstract}
We consider  the supersymmetric approach to gaussian 
disordered systems like the  random bond Ising model and  Dirac model
with random mass and random potential. These models appeared in particular
in the study of the integer quantum Hall transition. The supersymmetric
approach reveals an $osp(2/2)_1$ affine symmetry at the pure critical
point. A similar symmetry should hold at other fixed points.
We apply methods of conformal field theory to determine the conformal
weights at all levels. These weights can generically be negative because of 
non-unitarity. Constraints such as locality allow us to quantize the level $k$
and the conformal dimensions. This provides a class
of (possibly disordered) critical points in two spatial dimensions.
Solving the Knizhnik-Zamolodchikov equations
we obtain a set of four-point functions which exhibit a logarithmic dependence.
These functions are related to logarithmic operators. 
We show how all such features have a natural setting in the superalgebra
approach as long as gaussian disorder is concerned.
\end{abstract}
\vspace{5cm}
\noindent
\hspace{1cm} April 1996\hfill\\
\hspace*{1cm} SPHT-T96/040\hfill\\
\hspace*{1cm} hep-th/9605062
\thispagestyle{empty}

\newpage

\setcounter{page}{1}

\section{Introduction}\label{intro}

There is growing evidence that some disordered systems at criticality
share two unusual features: The existence of logarithmic operators
in the spectrum of the theory and the existence of an infinite number of
relevant operators with negative conformal dimensions. 
Logarithmic operators seem to be connected to hidden continuous symmetries.
In \cite{ckt} logarithmic operators were found to generate a change 
in a coupling constant of the effective WZNW action for $SU(2r)$, 
$r\rightarrow 0$, obtained after use of the replica trick. The existence 
of an infinite number of conformal operators \cite{mcw} means that there is
an infinite number of relevant perturbations which render the critical 
point unstable. Such a set of conformal dimensions is also related to 
the phenomenon of multifractality. 

Our aim in this paper is to show that both features follow naturally from
a supersymmetric treatment of the disorder. We consider  the 
two-dimensional
random bond Ising model at criticality and the random Dirac model in $2+1$
dimensions. These gaussian models 
allow a supersymmetric formulation of averages over disorder 
\cite{efetov}. One identifies a global superalgebra symmetry of the
effective action and assumes  it is enhanced
to an affine symmetry at a new 
critical point \cite{mcw,db}, as happened at the pure 
critical point.
One then derives the Sugawara stress-energy tensor
and obtains the set of conformal dimensions associated with the primary fields.
Additional constraints such as locality  allow
to further restrict the operator content of the theory. This provides
a class of, possibly disordered, critical points in two spatial dimensions. 
The specific structure of  the superalgebra implies that 
such dimensions  can  generically be negative. The presence of logarithmic
operators is also straightforward from the point of view of 
this algebra. Unlike ordinary Lie algebras, superalgebras  have
indecomposable (not fully reducible) representations. It is then 
possible to show on general grounds  that such representations
imply the existence of logarithmic operators and logarithms in correlation
functions. 

This paper is organized as follows. In section 2 we briefly recall the
supersymmetric approach for the random bond Ising model and the random Dirac
model. In section 3 we write down the $osp(2/2)_k$ current algebra and find
the Sugawara stress-tensor. In section 4 we briefly review the representations
of $osp(2/2)$ and obtain the conformal weights of primary fields. In section 5
we derive the quantization of $k$ from locality constraints. In section 6
we show how indecomposable representations lead to logarithmic
operators and consider an atypical representation. In section 7 
we obtain some logarithm-dependent four-point
functions. We conclude in section 8.

\section{The supersymmetric method for disordered systems}

The supersymmetric method is applicable to
models which are gaussian at fixed disorder, but it provides a
good starting point for disentangling properties of a large class
of disordered conformal field theories.
The study of the random bond Ising model reveals that an appropriate
algebraic framework for studying gaussian disordered
systems at criticality should be based on affine Lie
superalgebra with zero superdimension. In the case of
the random Ising model this algebra is $osp(2N/2N)$.
The fact that the algebra has an equal number of bosonic and fermionic
generators ensures that the Virasoro central charge
of the Sugawara stress-tensor vanishes. This is needed
by construction for a disordered system. The vanishing of the
central charge does not imply the triviality of the theory
since it is not unitary. 
The relevance of affine Lie superalgebras
was independently realized in refs. \cite{mcw,db}. 

We now briefly recall the supersymmetric method for the  study
of the random bond Ising model and the random Dirac model. 
These models have been  analyzed using the replica method in
refs. \cite{DoDo,Lu1,Shan}. The random bond Ising model
has also been studied in the context of massless scattering 
theories \cite{mussardo}.

\vskip 10pt
{\it The random bond Ising model}

In the scaling limit, near criticality, the Ising model is
described by a massive real Majorana fermion with mass
$\ov{m}\sim \tau = \frac{(T_C-T)}{T_C}$ were $T_C$ is the critical temperature.
In  the presence of disorder, {\it i.e.} when the coupling
constants between sites belong to a random set,
the mass becomes a function of space position.
The random Ising model is defined by the action $(z=x+iy)$:
\begin{eqnarray}
S[m(x)]=\int \frac{d^2x}{4\pi}
\left( \psi \partial_\zb \psi + \ov{\psi} \partial_z 
\ov{\psi} + i\, m (x) \ov{\psi} \psi \right)
\label{act1}
\end{eqnarray} 
where $\psi$ and $\ov{\psi}$ are grassmanian fields.
The mass $m(x)$ is chosen to be a quenched random 
variable with a gaussian measure:
\begin{eqnarray}      
P[m] = \exp\left({- \inv{4g} \int \frac{d^2x}{2\pi}
(m(x)-{\ov m})^2 }\right) \; . \label{mesur}
\end{eqnarray} 

The energy operator $\ep(x) = i\ov{\psi}(x)\psi(x)$  has 
dimension one. The Harris criterion
tells us that randomness in the bond interaction is
marginal in the 2d Ising model. It turns out that it is not
exactly marginal but only marginally irrelevant. 
At criticality the disorder only induces
logarithmic corrections to the pure system.

In order to compute averages of products of
correlation functions one introduces a
number of copies of fermions and of their supersymmetric
partners equal to the number of correlation 
functions in the product \cite{efetov,db}. 
One then rewrites these averages as fermionic
and bosonic path integrals. For the Ising model with $\ov{m}=0$
one obtains the following effective action  for the disorder 
average of the product of two correlation functions:
\begin{eqnarray}
S_{\rm eff} &=& \int \frac{d^2x}{2\pi}\left(
\psi_- \partial_\zb \psi_+ + \ov{\psi}_- \partial_z \ov{\psi}_+
+\eta\partial_\zb \ga + \etab\partial_z\gab\right)
+\frac{g}{8} \int\frac{d^2x}{\pi}~\Phi_{\rm pert} \nonumber\\
&=& S_* + \frac{g}{8} \int\frac{d^2x}{\pi}~\Phi_{\rm pert}
\label{Seff1}
\end{eqnarray}
with
\beq
\Phi_{\rm pert}=\left(\ov{\psi}_- \psi_+ - \psi_- \ov{\psi}_+
+\etab\ga-\eta\gab \right)^2 \label{pert1} \;.
\eeq
The $\psi$ are complex fermions and $\eta$ and $\gamma$ are complex
bosonic fields.
This action can be viewed as a perturbation of the (non-unitary)
conformal field theory specified by the action $S_*$.
This fixes the normalization of the fields to be:
\begin{eqnarray}      
\psi_-(z) \psi_+(w) \sim \inv{z-w} \quad,\quad
\ga(z)\eta(w) \sim \inv{z-w} \; . \label{norm}
\end{eqnarray} 
The central charge of the Virasoro algebra is zero. Note that since
the fermions $\psi_\pm$ have dimension one half the perturbing
field  $\Phi_{\rm pert}$ has dimension two.
It is therefore marginal.

The conformal field theory specified by $S_*$ is
invariant under an affine supersymmetric algebra whose
conserved currents are:
\bea
G_\pm(z) = \eta(z)\psi_\pm(z) \quad&,&\quad
\widehat G_\pm(z) = \ga(z)\psi_\pm(z)\; , \nonumber\\
K(z) = \eta^2(z) \quad&,&\quad \widehat K(z)=\ga^2(z)\; , \nonumber\\
J(z)=:\psi_-(z)\psi_+(z): \quad&,&\quad
H(z)=:\ga(z)\eta(z):\;\; .\label{curr}
\eea 
The dots refer to fermionic and bosonic normal ordering. 
There are four fermionic currents, $G_{\pm}$ and $\widehat G_{\pm}$,
which are generators of supersymmetric transformations, and four bosonic ones.
They form a representation of the affine $osp(2/2)$ \cite{snr,marcu}
current algebra at level one. 

The perturbing field  can also be written as a bilinear
in the currents,
\beq
\Phi_{\rm pert}=2\left[ \ov{J} J - \ov{H} H +\half (\ov{K} \widehat K
+\ov{\widehat K} K) + \ov{G}_- \widehat G_+ -
\ov{\widehat G}_- G_+ + \ov{G}_+ \widehat G_- - \ov{\widehat G}_+ G_-
\right]\,.
\eeq
In other words the perturbation (\ref{pert1}) is
a current-current perturbation. Therefore the
action $S_{\rm eff}$ preserves a global $osp(2/2)$ symmetry.

\vskip 10pt
{\it The random Dirac model}

The random  model of Dirac fermions
has been introduced in connection with
the quantum Hall transition \cite{luetal}.
Its action is:
\bea
S &=&\int \frac{d^2 x}{2\pi}\left(
  \psi_- \partial_\zb \psi_+ + \ov{\psi}_- \partial_z \ov{\psi}_+ 
\right.\nonumber\\
 &&\left. +i\frac{m(x)}{2} (\ov{\psi}_- \psi_+ -\psi_- \ov{\psi}_+ )
 +i\frac{V (x)}{2} (\ov{\psi}_- \psi_+ +\psi_- \ov{\psi}_+ )\right)\; .
\eea

The random variables $m$ and $V$ have a
gaussian distribution with widths $g_M$ and $g_V$.
We denote by $\Phi_M$ and $\Phi_V$ the
perturbing fields coupled to the constants
$g_M$ and $g_V$ after averaging over the disorder.
In the two-copy sector we can write them in terms of
the $osp(2/2)$ currents:
\bea
\Phi_V &=& 2 \ov{H} H - 2 \ov{J} J + \ov{K} \widehat K + \ov{\widehat K} K  
+ 2 \ov{G}_- \widehat G_+ - 2 \ov{\widehat G}_+ G_-
+ 2\ov{\widehat G}_- G_+  -2 \ov{G}_+ \widehat G_- \; ,\\
\Phi_M &=& 2 \ov{J} J  - 2 \ov{H} H  + \ov{K} \widehat K + 
\ov{\widehat K} K  + 2 \ov{G}_- \widehat G_+ - 2 \ov{\widehat G}_+ G_-
- 2\ov{\widehat G}_- G_+  + 2 \ov{G}_+ \widehat G_- \; .
\eea
We see that $\Phi_M=\Phi_{\rm pert}$. 

Both $\Phi_M$ and $\Phi_V$ preserve a global $osp(2/2)$ symmetry, whose
generators are
\beq
\begin{tabular}{rrrr}
$H_0+\overline H_0\,,$  & $\quad J_0+\overline J_0\,,$ &
$\quad K_0+\overline K_0\,,$ &
$\quad {\widehat K}_0+\overline{\widehat K}_0\,,$  \\
$G_{+0}+\overline G_{+0}\,$ &  $\quad {\widehat G}_{+0}+
\overline{\widehat G}_{+0}\,,$ &
$\quad G_{-0}+\overline G_{-0}\,,$ & $ \quad {\widehat G}_{-0}+
\overline{\widehat G}_{-0}\,.$
\end{tabular}
\eeq
for the  perturbation $\Phi_M$, and
\beq
\begin{tabular}{rrrr}
$H_0+\overline H_0\,,$  & $\quad J_0+\overline J_0\,,$ &
$\quad K_0-\overline K_0\,,$ &
$\quad {\widehat K}_0-\overline{\widehat K}_0\,,$  \\
$G_{+0}+\overline G_{+0}\,,$ &  $\quad {\widehat G}_{+0}-
\overline{\widehat G}_{+0}\,,$ &
$\quad G_{-0}-\overline G_{-0}$\,, & $ \quad {\widehat G}_{-0}+
\overline{\widehat G}_{-0}\,.$
\end{tabular}
\eeq
for the perturbation $\Phi_V$. Here the index $0$ denotes the zero
modes of the currents $J^a(z)=\sum_n J^a_nz^{-n-1}$.
If both perturbations are present,
the interaction preserves an $u(1/1)$ symmetry generated by
\beq
H_0+\overline H_0\,, \quad J_0+\overline J_0\,,  \quad G_{+0}+
\overline G_{+0}\,, \quad
{\widehat G}_{-0}+\overline{\widehat G}_{-0}\,.
\eeq 

The perturbative study of the
Dirac theory with a random potential and a random mass
is very similar to the perturbative study
of the random mass Ising model.
However the crucial difference is that contrary to
the randomness of the mass, the randomness of the potential
is marginally relevant. 
At $g_M=0$, the one-loop beta function is given by $\dot g_V=8 g_V^2$.
This means that $g_V$ grows at large distances. The infrared fixed 
point has not yet been determined \cite{luetal}. 
Should the affine symmetry, present at the UV fixed point of the pure system,
be restored at the IR point, then we 
believe this point should belong to the set we find in 
section \ref{quantiz}.

\section{The current algebra approach}

The critical free theory described by the action $S_*$ has a current
algebra symmetry, $osp(2/2)_1$, on both its  holomorphic and antiholomorphic
sectors. When randomness is introduced in the pure system the effective action
is no longer critical, and the current algebra symmetry is reduced
to a global $osp(2/2)$ symmetry. However  at a new critical point the
conformal invariance restores the current algebra symmetry. 
The value of the level $k$ is not preserved by the renormalization
flow. Therefore its value at the IR fixed point could be different
from its value at the UV point. 
It is possible to extract
properties at the new critical point by studying the current algebra
and its associated stress-energy tensor. 
In this context we consider $osp(2/2)_k$ at arbitrary  $k$
and obtain the Sugawara tensor. 

\subsection{The $osp(2/2)_k$ algebra}

In this section we write down the singular terms of 
the Operator Product Expansions \cite{ginsparg} satisfied by the currents
of the affine $osp(2/2)$ algebra at level $k$. 
The non-trivial OPE of the currents
are easily obtained from the currents (\ref{curr}). The level $k$
appearing below is equal to 1 for these currents. We find:
\bea
J(z)J(w) \sim \frac{k}{(z-w)^2}
\quad &;& \quad
H(z)H(w) \sim \frac{-k}{(z-w)^2} \nonumber\\
J(z)G_\pm(w) \sim \frac{\pm1}{z-w} G_\pm(w)
\quad &;& \quad
J(z)\widehat G_\pm(w) \sim \frac{\pm1}{z-w} \widehat G_\pm(w)\nonumber\\
H(z)G_\pm(w) \sim \frac{1}{z-w} G_\pm(w)
\quad &;& \quad
H(z)\widehat G_\pm(w) \sim \frac{-1}{z-w} \widehat G_\pm(w)\label{opecurrent}\\
H(z) K(w) \sim \frac{2}{z-w} K(w)
\quad &;& \quad
H(z)\widehat K(w) \sim \frac{-2}{z-w}\widehat K(w)\nonumber\\
\widehat G_\pm(z) G_\mp(w) &\sim& \frac{k}{(z-w)^2}
+ \frac{1}{z-w}(H(w)\pm J(w)) \nonumber\\
\widehat K(z) K(w) &\sim& \frac{2k}{(z-w)^2}
+\frac{4}{z-w} H(w) \nonumber\\
G_-(z) G_+(w) \sim \frac{1}{z-w} K(w)
\quad &;& \quad
\widehat G_-(z) \widehat G_+(w) \sim \frac{1}{z-w} \widehat K(w) \nonumber\\
K(z) \widehat G_\pm(w) \sim \frac{-2}{z-w} G_\pm(w)
\quad &;& \quad
\widehat K(z) G_\pm(w) \sim \frac{2}{z-w} \widehat G_\pm(w)\nonumber
\eea

Requiring the algebra to be associative constrains the possible
central terms extensions to the ones appearing in the OPE's (\ref{opecurrent}).
We can rewrite these OPE's in a more compact form as
\beq
J^a(z) J^b(w) \sim k \frac{\kappa^{ab}}{(z-w)^2} +f^{ab}_{\;\;\;\; c}\;
\frac{J^c(w)}{z-w}\; .\label{jj}
\eeq
The $f^{ab}_{\;\;\;\; c}$ are the structure 
constants of the Lie superalgebra $osp(2/2)$,
and $\kappa^{ab}$ is proportional to its non-degenerate Killing form.

\subsection{The Sugawara stress-energy tensor}

We now construct the Sugawara stress-energy tensor \cite{suga}. 
It is bilinear
in the currents defined in the previous section. Because of singularities
which appear at coinciding points one has to consider a regularized
version where normal ordered products of currents appear. This
is equivalent to a point splitting procedure where the singular parts
appearing in the OPE's of the currents are subtracted.
We take the usual definition for the  normal ordered product
of two fields $A(z)$ and $B(w)$:
\beq
:AB:(w) \equiv \oint_w\frac{dz}{2\pi i}\frac{A(z)B(w)}{z-w}
\label{nord}
\eeq
It should be noted that this ordering prescription does not coincide in general
with the Wick prescription used in (\ref{curr}). 
The Sugawara stress-energy tensor $T(z)$ is given
by
\beq
T(z)= \frac{1}{\kappa} \kappa_{ab} :J^a(z) J^b(z): \label{suga}
\eeq
where $\kappa_{ab}$ is the inverse of  $\kappa^{ab}$.
The constant $\kappa$ is determined by requiring $J^a(z)$ to be
a primary field of conformal weight one:
\beq
T(z)J^a(w) \sim \frac{J^a(w)}{(z-w)^2} + \frac{\partial J^a(w)}{z-w}\;.
\label{tjprim}
\eeq
When calculating OPE's one has to add a minus sign each time two odd
generators are permuted.
We find
\bea
T(z) &=&\frac{1}{4-2k}\left( : H(z)H(z)-J(z)J(z)-\frac{1}{2}(K(z)\widehat{K}(z)
+\widehat{K}(z)K(z))\right. \nonumber\\
&+& \left. \widehat{G}_+(z)G_-(z) -G_-(z)\widehat{G}_+(z) + 
\widehat{G}_-(z)G_+(z) -G_+(z)\widehat{G}_-(z):\right) \;.
\eea
The Sugawara tensor also satisfies
\beq
T(z)T(w)\sim \frac{c/2}{(z-w)^4} + 2 \frac{T(w)}{(z-w)^2} +
\frac{\partial T(w)}{z-w} \;.
\eeq
For a general Lie superalgebra, it is easy to show that the Virasoro central 
charge $c$ is proportional to the
superdimension; this is the difference between the number
of even and odd generators. Therefore $c$ vanishes for
the $osp(2/2)$ algebra, and more generally for any superalgebra
with an equal number of even and odd generators.
More precisely one has:
\beq
c=2 k\,\frac{\kappa_{ab}\kappa^{ab}}{\kappa}= 
2 k \,\frac{{\rm sdim}\, G}{\kappa}=0 \; .
\eeq
Because $c$ vanishes
$T(z)$ is a primary field, contrary to the case $c\neq 0$ where $T$ is 
just a level-two descendant of the unit operator. 

\section{Primary Fields}

We briefly describe the representations of $osp(2/2)$ 
and find the corresponding conformal weights.

\subsection{Some $osp(2/2)$ representations \label{reps}}

Unlike ordinary Lie algebras, the are two types of representations
for most superalgebras. The {\em typical} representations are irreducible and 
are similar to those of ordinary Lie algebras.
The {\em atypical} representations have no counterpart
in the ordinary Lie algebra setting. They can be irreducible or not
fully reducible (read reducible but indecomposable).

The superalgebra $osp(2/2)$ is isomorphic to the superalgebra $spl(2/1)$.
The representation theory of the latter algebra was studied
in \cite{snr,marcu}.
The quadratic Casimir of $osp(2/2)$  is
\beq
C_2 =\half\left( H^2 -\half (K\widehat K+\widehat K K)-J^2+
\widehat G_+G_--G_-\widehat G_++\widehat G_-G_+-G_+\widehat G_-\right)\;.
\eeq
The four even generators $K,\widehat K,H,J$ form a
$su(2)\oplus u(1)$ subalgebra.
The correspondence with the notation of \cite{snr}
is: $Q_+=-\widehat K/2$, $Q_-=K/2$, $Q_3=-H/2$, $B=J/2$,
$V_+=\widehat G_+/\sqrt{2}$, $V_-=-G_+/\sqrt{2}$, $W_+=-\widehat G_-/\sqrt{2}$,
$W_-=G_-/\sqrt{2}$. 

Let $b$ and $q_3$ be the eigenvalues of $B$ and $Q_3$.
Generically, a representation $(b,q)$,
$b \in \bC  \; ,\; q=0,\half,1,\tde,...$,
contains four $su(2)\oplus u(1)$ multiplets:
\bea
|b,q,q_3\rangle \; &,&\; q_3=-q,-q+1,...,q-1,q \;\; {\rm if}\;\; q\geq 0 \;,
\label{rep1}  \\
|b+\half,q-\half,q_3\rangle \; &,& \;q_3=-q+\half,...,q-\tde,q-\half 
\;\; {\rm if}\; \; q\geq \half\;, \label{rep2} \\
|b-\half,q-\half,q_3\rangle \; &,& \;q_3=-q+\half,...,q-\tde,q-\half
\;\;{\rm if}\;\; q\geq \half\;,\label{rep3} \\
|b,q-1,q_3\rangle \; &,& \; q_3=-q+1,...,q-2,q-1 \;\;{\rm if}\;\; q\geq 1\; .
\label{rep4}
\eea
The action of the four even generators on these multiplets is the one implied
by the notation. The four odd generators mix the different multiplets. 
The vector $|v\rangle=|b,q,q\rangle$ is a highest weight
vector, {\it i.e.} it satisfies
\beq
\widehat K |v\rangle=\widehat G_+ |v\rangle=\widehat G_- |v\rangle=0\;.
\eeq
The quadratic Casimir of this representation is $C_2=2(q^2-b^2)$.

If $b\neq \pm q$ the representation is denoted by $[b,q]$ and is typical;
the quadratic and cubic Casimirs do not vanish. All the vectors in 
the representation can be obtained from the highest weight 
vector $|v\rangle$ by applying on it polynomials in the generators.
The representation $[b,q]$ has dimension $8q$. 
The  representation $[0,1/2]$ is four-dimensional
and contains one spin $1/2$ and two spin $0$ multiplets.

{\it Atypical representations}

When $b=\pm q$ several kinds of atypical representations arise.
Both Casimirs vanish, and yet these representations are not the
trivial one-dimensional representation.
One kind has dimension $4q+1$ and is denoted by $[q]_{\pm}$.
To obtain $[q]_+$ (resp. $[q]_-$)
one drops the two multiplets
(\ref{rep3}) and (\ref{rep4}) (resp. (\ref{rep2}) and (\ref{rep4})).
These representations are irreducible.

{\it Atypical indecomposable representations}

Generally, they are semidirect sums of atypical irreducible
representations. They can contain two, three or four terms
and they  arise in tensor products of irreducible
representations.
An interesting example, containing four irreducible representations,
arises in the tensor product of two 
representations $[0,1/2]$. The result
is the direct sum of $[0,1]$ and
an eight-dimensional representation which is the semidirect sum
of $[1/2]_-$, $[1/2]_+$ and two $[0]$ representations.
We call it $[0,-1/2,1/2,0]$ (see fig. 1 b). 
The vector $s$ is
invariant; it is annihilated by all the generators. The quadratic Casimir
$C_2$ vanishes on all states except $t$: $C_2\ t=4 s$.
We have
\beas
s&=&\half \left( |\half,0\rangle\ot |-\half,0\rangle-|0,\half\rangle\ot
|0,-\half\rangle + |0,-\half\rangle\ot|0,\half\rangle + |-\half,0\rangle\ot
|\half,0\rangle \right) \\
t&=&\half \left( |\half,0\rangle\ot |-\half,0\rangle+|0,\half\rangle\ot
|0,-\half\rangle - |0,-\half\rangle\ot|0,\half\rangle + |-\half,0\rangle\ot
|\half,0\rangle \right)
\eeas

\subsection{Tensor product of $osp(2/2)$ representations}

The tensor product of two irreducible representations of a
superalgebra is not
necessarily completely reducible.
Ref. \cite{snr} gives a sufficient condition for a tensor 
product of two $osp(2/2)$ representations to be completely reducible,
and the irreducible components. The result is the following
\bea
\label{tpr}
&[b,q]\otimes [b',q']=[b+b',q+q']\oplus [b+b',q+q'-1]\oplus\ldots
\oplus [b+b',|q-q'|]& \nonumber \\
&\oplus [b+b',q+q'-1] \ldots \oplus [b+b',|q-q'|+1]&\nonumber \\
&\oplus [b+b'+1/2,q+q'-1/2] \ldots \oplus [b+b'+1/2,|q-q'|+1/2]&\nonumber \\
&\oplus [b+b'-1/2,q+q'-1/2] \ldots \oplus [b+b'-1/2,|q-q'|+1/2]& \\
& {\rm if} \;\; \pm b>q\geq \half, \quad \pm b'>q' \geq \half \; .&\nonumber
\eea
For the other values of $b$, $b'$, this expression still gives the
correct content of $su(2)\oplus u(1)$ charges of the tensor product.
This expression also permits to obtain some information on the 
reducibility: if two components have different Casimirs, 
then they belong to different (maybe indecomposable) representations.
Vanishing Casimirs are, generally, signs of some pathologies.

\vskip 0.5cm
\epsfxsize 16.0 truecm
\epsfbox{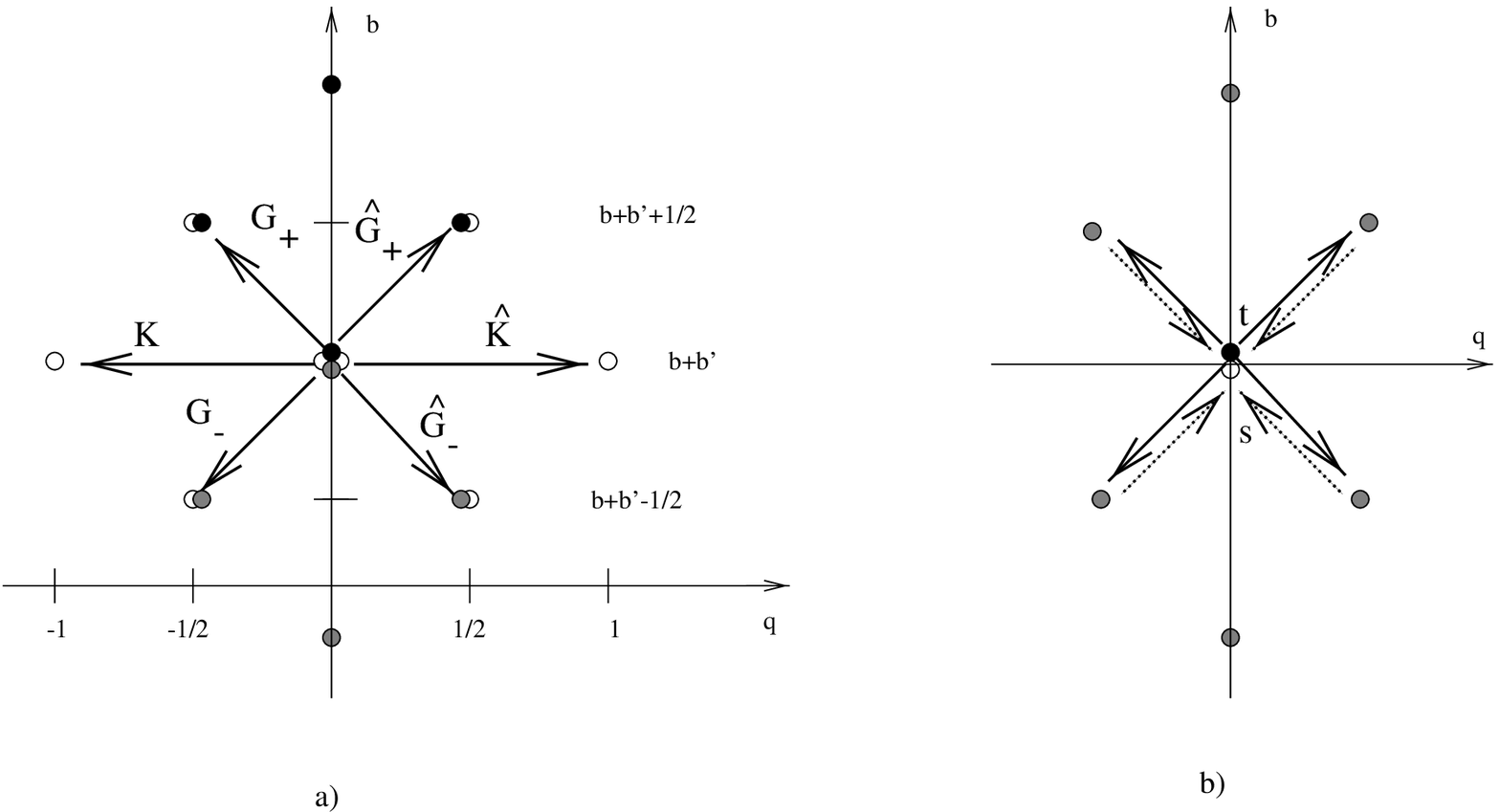}
\vskip 0.5cm
{\small {\bf Fig.1:} a) The tensor product of the representations
$[b,1/2]$ and $[b',1/2]$. If $b>1/2$, $b'>1/2$, the result is the direct
sum of $[b+b',1]$ (white dots), $[b+b'+1/2,1/2]$ (black dots) and
$[b+b'-1/2,1/2]$ (grey dots). The arrows represent the action of the 
generators.
 b) The representation $[0,-1/2,1/2,0]$.
The vector $t$ (black dot) is cyclic and $s$ (white dot) is invariant.
The action of the fermionic generators $\widehat G_+$,
 $\widehat G_-$, $G_+$ and  $G_-$
is represented by arrows.}
\vskip 0.4cm

\subsection{$osp(2N/2N)$ representations}

When calculating the disorder average of product of correlation 
functions for the Ising and Dirac models, we can introduce an 
arbitrary number of copies of
fermions and bosons. One is led is to consider $osp(2N/2N)$ as symmetry. 
The theory is consistent if the results
of the calculations do not depend on $N$.
One step in this direction is to show that the conformal dimensions of the
primary fields are independent of $N$.
 
The main difference between the superalgebras $osp(2/2)$ and $osp(2N/2N)$
with $N>1$ is that the even subalgebra of the first one is not simple.
This is why representations of the former are indexed by a 
continuous parameter, while the representation of the 
latter are characterized only by discrete variables.
According to the classification of the superalgebras \cite{kac}, $osp(2/2)$
is the first of the series $C(N+1)$, while $osp(2N/2N)$ is a superalgebra
of the type $D(N,N)$.
 
The finite dimensional typical representation of the superalgebras were
classified by Kac \cite{kac}.
For $D(N,N)$, the typical representations are characterized by
their highest weights, $\Lambda=\sum_{i=1}^Na_i\delta_i+\sum_{i=1}^Na_{N+i}
\epsilon_i\,;$ $\delta_i$ and $\epsilon_i\,,$ $i=1,N$ form
an orthogonal basis with $(\epsilon_i,\epsilon_j)=-\delta_{ij}\,,$
$(\delta_i,\delta_j)=\delta_{ij}\,,$ $(\epsilon_i,\delta_j)=0$.
The `numerical marks' $a_i$ satisfy the following conditions:
 
i) $a_i \in \bZ_+$, $i\neq N$,
 
ii) $j=a_N-a_{N+1}-...-a_{2N-2}-\half(a_{2N-1}+a_{2N}) \in \bZ_+$,

iii) $a_{N+j+1}=...=a_{2N}=0$, if $j\leq N-2$; $a_{2N-1}=a_{2N}$,
if $j=N-1$.

The first two conditions express the fact that $\Lambda$ is a
dominant weight for the even algebra $Sp(2N)\ot O(2N)$.

The value of the quadratic Casimir for the representation with
highest weight $\Lambda$ is given, up to a normalization,
by $(\Lambda, \Lambda+2\rho)$, with $\rho=\rho_0-\rho_1$ and $\rho_{0(1)}$
is half the sum of the even (odd) positive roots.
Considering the positive roots associated to the distinguished
Dynkin diagram (the Dynkin diagram with a single fermionic root),
we obtain
$$\rho=\sum_{i=1}^N (N-i)(\epsilon_i-\delta_{N-i+1})\;.$$
We normalize the value of the quadratic Casimir of the adjoint
representation $(2,0,...,0)$ to be $1$; the Casimir of the representation
$(a_1,...,a_{2N})$ will be
$$C_2(\Lambda)={1\over 4}\sum_{i=1}^N
\left[(a_i-i+1)^2-(a_{N+i}+N-i)^2\right]\;.$$
For the fundamental representation $(1,0,...,0)$ we have $C_2=1/4$,
independent of $N$.
 
In order to obtain compatibility between the results obtained in the
$osp(2/2)$ and $osp(2N/2N)$ frameworks, the continous parameter $b$
of the $osp(2/2)$ representations has to be constrained to the discrete 
values compatible with the value of the discrete parameters of the 
$osp(2N/2N)$ representations.

\subsection{Conformal dimensions of the primary fields}

The primary fields $\phi^\af(w)$ are highest weight vectors
of affine $osp(2/2)_k$ and of the Virasoro algebra.
In terms of operator products one has:
\bea
J^a(z) \phi^\be(w) &\sim& (T^a)^{\ga\be}\;\frac{\phi^\ga(w)}{z-w}\;,
\label{jprim}\\
T(z) \phi^\af(w) &\sim& \Delta\frac{\phi^\af(w)}{(z-w)^2}
+\frac{\partial \phi^\af(w)}{(z-w)}\;.\label{tprim}
\eea
Using (\ref{jj}) and (\ref{jprim}) one then shows that the matrices 
$T^a$ form a representation of the
$osp(2/2)$ algebra. But this applies to any algebra. 
Using the Sugawara form (\ref{suga}) of the stress-energy tensor
and (\ref{jj}) one obtains the conformal weights corresponding to
a particular representation. For the representations
$[b,q]$ and $[q]_{\pm}$ the Casimir $C_2$ is diagonal and 
we obtain:
\beq
\De_{(b,q)}=\frac{1}{\kappa} \kappa_{ab}T^a T^b=\frac{2(q^2-b^2)}{2-k}\;.
\eeq

We are also able to calculate the conformal dimensions of the primary
fields for all $N$. Due to the fact that the dual Coxeter number of
$osp(2N/2N)$ equals 1 for all $N$, these conformal dimensions do
not change if we change the number of copies to $N'>N$; for example
$$\De_{(1,0,...,0)}={1\over 4-2k}$$
is the conformal dimension of the fields in the fundamental representation.
                                
The conformal dimensions are potentially negative;
this reflects the non-unitarity of the theory we considered.

\section{Local Operators and Quantization of $k$ \label{quantiz}}

As seen in section two the interactions generated by
the random mass and random scalar potential conserve diagonal
$osp(2/2)$, $J_x^a=J_0^a\pm\bar J_0^a$. We expect to find
in the spectrum of the theory, at the new fixed point for the
potential interaction, particles
which carry the same charges $H_x$ and $J_x$ as the original ones (bosons and
fermions).  The fields corresponding to these particles are
no more  holomorphic or antiholomorphic, but they can be made
up of the new primary fields.
Suppose that they are of the form
\beq
\label{champs}
\phi(z,\bar z)=\phi_{[b,q]}(z)\phi_{[b',q']}(\bar z)
\eeq
where $\phi_{[b,q]}(z)$ ($\phi_{[b',q']}(\bar z)$) are primary
fields belonging to the representation
$[b,q]$, respectively $[b',q']$ of the two chiral algebras $J^a(z)$
and $\bar J^a(\zb)$.

The charges of the bosons and fermions with respect to the currents
$J^a_x$ are:
\beq
\label{fb}
\begin{tabular}{rrr}
$\  $&$J_x $ &$ H_x$ \\
$\psi_+(z) \qquad $ & $1/2$ &$0$ \\
$\psi_-(z) \qquad$  & $-1/2$ &$0 $ \\
$\gamma(z) \qquad$  & $0$ &$1/2 $ \\
$\eta(z) \qquad$  & $0$ &$-1/2 $
\end{tabular}
\eeq
We have to identify the representation $[0,1/2]$
in the tensor product of the
representations $[b,q]$ and $[b',q']$.
Using equation (\ref{tpr}) we obtain the following possibilities
\beq
\label{spinb}
 a) \quad q=q', \quad b+b'\pm 1/2=0\;,\;\;\;\;
b) \quad |q-q'|=1/2, \quad b+b'=0\;.
\eeq

As in  \cite{mcw}, in order to obtain a sensible theory,
we impose the constraint of locality
on the fields $\phi(z,\bar z)$.
The OPE of two such composite fields can be written as
\beq
\label{ope}
\CO_a(z,\zb) \CO_b(0,0)=\sum_c C_{ab}^c\, z^{\De_c-\De_a-\De_b}
\ \zb^{\bar\De_c-\bar\De_a-\bar\De_b} \CO_c(0,0)\;.
\eeq
The operators $\CO_a$ and $\CO_b$ are said to be mutually local
if the OPE is single-valued, which means that the spin difference
is an integer
\beq
s_c-s_a-s_b=\De_c-\De_a-\De_b-\bar \De_c+\bar \De_a+ \bar \De_b
\in \bZ\;.
\eeq
Let us take $\CO_a(z,\zb)=\CO_b(z,\zb )
=\phi_{[b,q]}(z)\phi_{[b',q']}(\zb)$. We evaluate the left and right
chirality OPE's separately. The
operators appearing in the OPE for the left chirality are the
operators associated to the representations $\af$ 
appearing in the tensor product
$[b,q]\otimes[b,q]$ (see eq. (\ref{tpr}))
\begin{eqnarray*}
\phi_{[b,q]}(z)\phi_{[b,q]}(0)
& \sim & \sum_{\af} z^{\De_{\af}-2\De_{[b,q]}}\; \phi_{\af}(0) 
\end{eqnarray*}
and similarly for the right chirality, with $[b,q]$ replaced by
$[b',q']$.
We ignored the possible existence of logarithms in this 
OPE (when two conformal dimensions are degenerate). 
We also supposed that all the primary fields allowed by the 
$osp(2/2)$ selection rules (and their 
descendants) appear in
the OPE. For a Kac-Moody algebra
this is not always the case; the operator content of the
theory depends on the value of the level $k$.
In fact the calculation shows that the constraint of locality
depends very weakly on how many of the operators $\phi_{\af}$
really appear in the OPE. 

The condition $s_c-s_a-s_b \in \bZ$ is fulfilled for all
the possible operators $\CO_c(0,0)$ in (\ref{ope}) if 
\bea
\label{quant}
&2l(4q-1)\in \bZ\;, \quad 2l(4q-2)\in \bZ\;, \quad 4l(q-b)\in \bZ\;,\\
&2l(4q'-1)\in \bZ\;, \quad 2l(4q'-2)\in \bZ\;, \quad 4l(q'-b')\in \bZ\;, \\
&{\rm and} \quad\;\; 4l(q^2-b^2-q'^2+b'^2) \in \bZ\;,
\eea
where $l=1/(2-k)$.
With the two possibilities in (\ref{spinb}) we obtain the following
restriction on $k$:
\beq
k=2-\frac{1}{l}\;, \qquad l\in \bZ\;. \label{restri}
\eeq

\section{Logarithmic operators \label{loga}} 

We argue that the appearance of  logarithmic operators is a
general feature of theories with an underlying superalgebra symmetry,
as long as indecomposable representations are concerned.
We give as an example the occurrence of
logarithmic correlations for $osp(2/2)$.

The first logarithmic correlation functions
in conformal field theory were obtained by Rozansky and Saleur in \cite{saleur}
in the framework of the $GL(1,1)$ WZW model. Later
Gurarie \cite{gurarie} pointed out that the logarithms are related
to non-diagonalizable Virasoro generators. See also \cite{ckt} for disordered
models,  \cite{kausch},  and 
\cite{flohr} in relation to the fractional quantum Hall effect. 
Suppose we are given two operators
$A_1(z)$, $A_2(z)$ with:
\bea 
T(z)A_1(0) &\sim & \frac{\De}{z^2}A_1(0)
+\frac{\partial A_1(0)}{z}\;,\label{ta1}\\
T(z)A_2(0) &\sim & \frac{\De}{z^2}A_2(0)+
\frac{\partial A_2(0)}{z}+\frac{a}{z^2}A_1(0)\;.\label{ta2}
\eea 
Then they transform under the dilatations $z'=\af\, z$ as
\beq 
A'_1(z')=\af^{-\De}A_1(z)\;,\;\;
A'_2(z')=\af^{-\De}A_2(z)+a\log\af \; A_1(z)\;.
\eeq
And more generally one has:
\bea
&\lbrack L_n,A_1(z)\rbrack =(z^{n+1}\partial_z+\De(n+1)z^n)A_1(z) \;, \\
&\lbrack L_n,A_2(z)\rbrack =(z^{n+1}\partial_z
+\De(n+1)z^n)A_2(z)+ a(n+1)z^n A_1(z)\;.
\eea
This gives rise to differential equations for the   
correlation functions of the operators $A_1(z)$ and $A_2(z)$, through the
conformal Ward identities 
for $n=-1,0,1$. Let us determine the two-point functions of these 
operators. The $n=0$ differential equations for these 
correlation functions are
\beas
&(z\partial_z+2\De)\vev{A_1(z)A_1(0)}=0\;, \;\;
(z\partial_z+2\De)\vev{A_2(z)A_1(0)}+a\vev{A_1(z)A_1(0)}=0\;,\\
&(z\partial_z+2\De)\vev{A_2(z)A_2(0)}+a\vev{A_1(z)A_2(0)}+
a\vev{A_2(z)A_1(0)}=0\;,
\eeas
while for $n=1$ we have
\beas
&(z\partial_z+2\De)\vev{A_2(z)A_1(0)}+2 a\vev{A_1(z)A_1(0)}=0\;,\\
&(z\partial_z+2\De)\vev{A_2(z)A_2(0)}+ 2 a \vev{A_1(z)A_2(0)}=0\;.
\eeas
This implies 
\beas
&\vev{A_1(z)A_1(0)}=0\,, \;\vev{A_2(z)A_1(0)}=\vev{A_1(z)A_2(0)}=
C z^{-2\Delta}\,,\\
&\vev{A_2(z)A_2(0)}= (-2a C\log z+C')z^{-2\De}\,,
\eeas
with $C$, $C'$ arbitrary constants. This system of correlation functions
appeared in \cite{gurarie,ckt}. 
Thus if the operators $A_1$, $A_2$ satisfy the OPE (\ref{ta1}--\ref{ta2})
they can have logarithmic correlators, and $A_2(z)$ is the logarithmic operator
related to $A_1(z)$. 

In the following we give an example of such operators 
appearing in the indecomposable representations of the $osp(2/2)$
superalgebra. 
Consider fields $\phi_i(z)$ in the atypical indecomposable
representation $[0,-1/2,1/2,0]$ (fig. 1b). 
The  structure of the generators $T^a$ is
\beq
T^a= \begin{array}{c}
     \\  \la \\ \la' \\ s \\ t
     \end{array}  
\begin{array}{c} \begin{array}{cccc}\la & \la' &s & t \end{array} \\
      \left(\begin{array}{cccc}
            t^a & 0 &0 &*\\
            0&t'^a& 0& * \\
            ** & * & 0& 0 \\
            0 & 0 & 0& 0
       \end{array} \right)
     \end{array}  
\eeq 
where $\la=[1/2]_+$, $\la'=[1/2]_-$ and the stars stand
for possible non-zero matrix elements.
The action of the currents on the fields
takes the form
\beq
J^a(z) \phi_1(0) = T^a_1 \frac{\phi_1(0)}{z}\;,\quad
J^a(z) \phi_2(0) = T^a_1 \frac{\phi_2(0)}{z} + T^a_2 \frac{\phi_1(0)}{z}\;,
\eeq
where the index 1 refers to the invariant subspace. This form is typical
of indecomposable representations of any algebra. We find
the most singular term arising
from the $J-\phi$ contractions in  $\vev{J^a(z)J^b(w)\phi_2(0)\prod{\cal O}}$.
Using the definition
\beq
T(z)=\lim_{z\rightarrow w}\frac{1}{\kappa} \left( \kappa_{ab} J^a(z)J^b(w)
-k \frac{\kappa_{ab}\kappa^{ab}}{(z-w)^2}\right)
\eeq
we conclude that 
\bea
T(z)\phi_1(0) &=& \frac{\Delta}{z^2} \phi_1(0) 
+ \frac{1}{z}\partial\phi_1(0)\;,\\
T(z)\phi_2(0) &=& \frac{\Delta}{z^2} \phi_2(0)
+ \frac{1}{\kappa}\frac{\kappa_{ab}(T^a_1 T^b_2+T^a_2 T^b_1)}{z^2} \phi_1(0)
+\frac{1}{z}\partial \phi_2(0) \;,
\eea
where $\Delta=\frac{1}{\kappa}\kappa_{ab} T^a_1 T^b_1$. 
This non-diagonal action of the Virasoro algebra is the direct
consequence of the existence of an indecomposable representation
at the affine level.  For the representation at hand one gets
$C_2\, t \sim s$, $C_2\, s=0$, and for the fields
$\phi_s$ and $\phi_t$ with zero $osp(2/2)$ charges:
\beq
T(z)\phi_s(0) \sim \frac{\partial \phi_s(0)}{z}\;,\;\;\;\;
T(z)\phi_t(0) \sim  a\frac{\phi_s(0)}{z^2}
+\frac{\partial \phi_t(0)}{z}\;.
\eeq
According to the previous discussion their correlation functions are
$$\vev{\phi_s(z)\phi_s(0)}=0\;,\;\;
\vev{\phi_t(z)\phi_s(0)}= C\;,\;\; \vev{\phi_t(z)\phi_t(0)}=
-2a C \log z + C'\;.$$
We can also solve the differential equations for
$\vev{\phi_s}$ and $\vev{\phi_t}$ to find $\vev{\phi_s}=0$, $\vev{\phi_t}=c$.

The indecomposable representation we considered
appear in the tensor product of two fundamental representations.
Thus logarithmic terms are expected to appear in the four 
point functions
(if the logarithmic operators are not identically 
zero, as seems to be the case for the free bosons/fermions $k=1$).

\section{Four-point functions}

We now calculate  a set of four-point correlation functions for 
primary fields in the $[0,\half]$ representation of $osp(2/2)$.
As explained in section \ref{loga} we expect to find a logarithmic
dependence.

\subsection{Differential equations} 

Consider the basis
\beq
|1\rangle =|0,\hf\rangle \; ,\; |2\rangle =|-\hf,0\rangle \; ,\; 
|3\rangle =|\hf,0\rangle \; ,\; |4\rangle =|0,-\hf\rangle \; ,
\label{basis}
\eeq
where $|1\rangle$ and $|4\rangle$ are even (bosonic) and $|2\rangle$ and 
$|3\rangle$ are odd (fermionic). These vectors are labeled 
by the eigenvalues of $B$ and $Q_3$. The odd generators are given by:
\bea
\label{vrep1}
V_+=\sqrt{2}\left( \begin{array}{cccc}
0 & \ep & 0 & 0 \\
0 & 0 & 0 & 0 \\
0 & 0 & 0 & \af \\
0 & 0 & 0 & 0    \end{array} \right) &,&
W_-=\sqrt{2}\left( \begin{array}{cccc}
0 & 0 & 0 & 0 \\
-\be & 0 & 0 & 0 \\
0 & 0 & 0 & 0 \\
0 & 0 & \ga & 0  \end{array} \right)\;, \\
\label{vrep2}
W_+=\sqrt{2}\left( \begin{array}{cccc}
0 & 0 & \ga & 0 \\
0 & 0 & 0 & \be \\
0 & 0 & 0 & 0 \\
0 & 0 & 0 & 0  \end{array} \right) &,&
V_-=\sqrt{2}\left( \begin{array}{cccc}
0 & 0 & 0 & 0 \\
0 & 0 & 0 & 0 \\
-\af & 0 & 0 & 0 \\
0 & \ep & 0 & 0  \end{array} \right) \;.
\eea
The four parameters appearing in (\ref{vrep1}--\ref{vrep2}) are constrained
by
\beq
4 \af \ga =1 \;\; ,\;\;\; 4 \be \ep = 1 \; .
\label{const}
\eeq
Thus there are two free parameters which correspond to arbitrary relative
normalizations of the  $su(2)$ doublet $(|1\rangle ,|4\rangle )$,
and the two singlets $|2\rangle$ and $|3\rangle$.

The vacuum invariance
under the zero modes of the affine algebra, or equivalently invariance under
global gauge transformations, imply the following linear equations
\beq
\sum_{i=1}^4 T^a_i \langle 
\phi_1(z_1) \phi_2(z_2)\phi_3(z_3)\phi_4(z_4)\rangle =0 \;.\label{glo}
\eeq
It is understood that minus signs arise each time we permute two fermionic
objects, and omit the antiholomorphic dependence and indices of the fields.
Similar equations hold for the `right' (antiholomorphic) chiral generators. 
These equations are solved in terms of invariant tensors for the 
representations at hand. We have  found the explicit expressions for three 
linearly independent tensors which we give in the appendix.

Requiring the  
fields of a specific module to be both affine and Virasoro primary 
gives a set of equations relating the fields. Such equations give partial 
differential equations, the Knizhnik-Zamolodchikov equations \cite{kz},
for the correlation functions.
One gets
\beq
\left[\half\frac{\partial}{\partial z_1} - \frac{\kappa_{ab}}{\kappa}
\sum_{j=2}^4 \frac{1}{z_1-z_j}T_1^a T_j^b \right] \langle \phi_1(z_1) 
\prod_{j=2}^4 \phi_j (z_j)\rangle =0 \; .
\eeq
After a conformal transformation this equation  reduces to a matrix 
differential equation for the correlators
\beq
F^{i_1 i_2 i_3 i_4}(z)=\lim_{w\rightarrow \infty} w^{2\De_4}
\langle \phi^{i_1}(z) \phi^{i_2}(0)\phi^{i_3}(1)\phi^{i_4}(w)\rangle\;.
\eeq
One gets
\beq
\half\partial F(z)= \frac{1}{\kappa} \left(\frac{1}{z} {\cal P} +\frac{1}{z-1}
{\cal Q}\right) F(z)\;,
\eeq
where
\beq
z=\frac{z_{12}z_{34}}{z_{23}z_{41}}\;,\;\;
{\cal P}=\kappa_{ab} T^a_1\otimes T^b_2\;,\;\;
{\cal Q}=\kappa_{ab} T^a_1\otimes T^b_3\;.
\label{kz}
\eeq
Similar equations hold for the antiholomorphic sector. 
The complete correlator is given by
\beq
{\cal F}(z_1,z_2,z_3,z_4)= z_{14}^{-2\De_1} z_{23}^{-\De_1-\De_2-\De_3+\De_4}
z_{24}^{\De_1-\De_2+\De_3-\De_4} z_{34}^{\De_1+\De_2-\De_3-\De_4} 
F(z) \;.
\eeq
For the foregoing representations $\Delta_i=1/(2 x)$ and 
the $z_{ij}$ prefactor reduces to $z_{14}^{-x^{-1}} z_{23}^{-x^{-1}}$,
where 
$$x=2-k\, .$$

The correlators have the $osp(2/2)\times osp(2/2)$ invariant decomposition
\beq
F^{i_1i_2i_3i_4,j_1j_2j_3j_4}(z,\zb)=
\sum_{A,B=1}^{3} I_A^{i_1i_2i_3i_4} \bar{I}_B^{j_1j_2j_3j_4} F_{AB}(z,\zb).
\eeq
The tensors are given in the appendix and the nine scalar functions satisfy
the differential equations:
\beq
x\frac{\partial F}{\partial z} =
\left[\frac{1}{z} P +\frac{1}{z-1} Q\right] F\;,
\eeq
where $F$ now denotes the vector $(F_{1B},F_{2B},F_{3B})$ 
for all $B$.
There are similar equations for the antiholomorphic dependence, for the vector
$(F_{A1},F_{A2},F_{A3})$. 
We give the matrices $P$ and $Q$ in the appendix.
Let $f_A(z)=F_{AB}(z)$ for any given $B$. It is straightforward  
to reduce this first-order matrix differential equation
to the following set of equations:\footnote{It is amusing to note that
the third-order differential equations for $f_1(z)$ and $f_2(z)$
reveal two {\em apparent} singularities at $z=2$ and $z=\half$.}
\bea
&x^3 z^3 (1-z)^3 \partial^3 f_3(z) + x^2 (1+2 x) z^2 (1-z)^2(1-2 z)\partial^2
f_3(z) +x z(1-z)[-1-x& \nonumber\\
& +2 x z-2 x(2+x)z(1-z)]\partial f_3(z) 
+(-x -1 +2 z+2 x z(1-z))f_3(z) = 0&
\eea
\bea
f_2(z) &=& -\frac{1}{4\ep\ga x z(1-z)}(x^2 D^2 f_3(z) +2x(1-z) D f_3(z) 
+(1-2 z)f_3(z) )\\
f_1(z)&=&\frac{1}{4\ep\ga} (x Df_3(z) -f_3(z) ) +(z-2) f_2(z)
\eea
where $D=z(1-z)\partial $.
As expected the equation
for $f_3$ is fuchsian with three regular singularities at $0$, $1$ 
and $\infty$.\footnote{See \cite{ince} or any book on differential equations.}
The corresponding indices are $(-x^{-1},-x^{-1},1+x^{-1})$, 
$(-x^{-1},1-x^{-1},x^{-1})$ and $(0,0,1+ 2 x^{-1})$. The degeneracy of some
indices, {\it i.e.} when the difference of two indices belongs to $\bZ$,
signals the eventual presence of logarithmic solutions. Thus one is naturally
lead to distinguish between different values of $x$'s. 

We were able to express the functions $f_i$ in terms of generalized 
hypergeometric functions. 
We obtained:
\bea
&F^{1423}(z)=f_3(z)= \hf (1-x\,D)F^-(z)\;,& \nonumber\\
&x^3 z^3(1-z)^3\partial^3 F^- + x^2(3x+1)z^2(1-z)^2(1-2z)
\partial^2 F^-&\nonumber\\
&+ z(1-z)(x^2(x+1)(1-6z(1-z))-x)\partial F^- -(1-2z)F^- =0\;.&\label{de1}
\eea

Equation (\ref{de1}) is fuchsian with three regular singular points at
0, 1 and $\infty$. The indices at 0 and 1 are $(-\frac{1}{x},-\frac{1}{x},  
\frac{1}{x})$. The  solutions of equation (\ref{de1}) can be written as 
generalized hypergeometric functions:
\bea
F^-_0(z) &=& (z(1-z))^{-\frac{1}{x}}\; {}_3F_2\left(\frac{1}{2},-\frac{1}{x},
1-\frac{1}{x};1,-\frac{2}{x};4z(1-z)\right)\\
F^-_1(z) &=& F^-_0(z) \log[4z(1-z)] + (z(1-z))^{-\frac{1}{x}}\sum_{n\geq 0}
c_n (4z(1-z))^n\\ 
F^-_2(z) &=& (z(1-z))^{\frac{1}{x}} \; {}_3F_2\left(\frac{1}{x},1+\frac{1}{x},
\frac{1}{2}+\frac{2}{x};1+\frac{2}{x},1+\frac{2}{x};4z(1-z)\right)
\eea
where 
\beas
& & {}_3F_2(a_1,a_2,a_3;b_1,b_2;z) \equiv \sum_{n\geq 0}\frac{(a_1)_n (a_2)_n
(a_3)_n}{(b_1)_n (b_2)_n n!} z^n\;,\;\; (a)_n = a(a+1)\cdots (a+n-1)\;, \\
&c_n& =\frac{d}{ds}
\left(\frac{\Gamma(n+s+\hf)\Gamma(n+s-\frac{1}{x})\Gamma(n+s+1-\frac{1}{x})
\Gamma(s+1)^2\Gamma(s+1-\frac{2}{x})}
{\Gamma(n+s+1)^2\Gamma(n+s+1-\frac{2}{x})\Gamma(s+\hf)\Gamma(s-\frac{1}{x})
\Gamma(s+1-\frac{1}{x})}\right)_{s=0}\;.
\eeas
It must be noted that although the solutions look symmetric under
the exchange $z\leftrightarrow 1-z$ it is {\em not}
possible to have three linearly independent solutions with such a symmetry.
When the series defining the hypergeometric functions do not truncate there
are singularities at $u= 4z(1-z) =1$ of the type $\log(1+\sqrt{1-u})$.
And since the reciprocal image 
of the interior of the $u$-unit disc consists of two disjoint lobes, in the
$z$ plane,  centered
around 0 and 1, the change $z \rightarrow 1-z$ changes the determination
of the hypergeometric function around $z=1$  from what it is around
$z=0$. This feature can be seen explicitly 
for the cases $x=1$ and $x=1/2$ below. 

Before analyzing some specific solutions let us note that after the
change of function $F^-(z)=(z(1-z))^{-\frac{1}{x}} G(z)$
one obtains
\bea
&z^2 (z-1)^2 \partial^3 G + z(1-z)(K_1 z+K_2 (z-1))\partial^2 G
+(L_1 z^2+L_2 (z-1)^2&\nonumber\\
&+L_3 z(z-1))\partial G + (M_1 z + M_2 (z-1))G=0\;,&
\eea
where
\beq
K_1=K_2=3-\frac{2}{x}\;,\;\; L_1=L_2=1-\frac{2}{x}\;,\;\; 
L_3=4(1-\frac{1}{x})^2\;,\;\; M_1=M_2=\frac{2}{x}(\frac{1}{x}-1)\;.
\eeq
Equations of this type appeared for correlations functions containing 
a level three null vector in minimal models (see ref.~\cite{dotfat}).
There are integral representations for the solutions 
\beq
G(z)=\int_{C_1}dt_1 \int_{C_2}dt_2 t_1^{\frac{1}{x}-1}(t_1-1)^{\frac{1}{x}-1}
(t_1-z)^{\frac{1}{x}} t_2^{\frac{1}{x}-1}(t_2-1)^{\frac{1}{x}-1} 
(t_2-z)^{\frac{1}{x}} (t_1-t_2)^{-\frac{4}{x}} \;.
\eeq
These integrals can however diverge, depending on the contours $C_1$ and $C_2$
and the value of $x$. 

\subsection{The level 1 correlators}

For $x=1$, {\it i.e.} $k=1$, we are able to write the solutions of 
(\ref{de1}) in a closed form:
\beq
F^-(z)=g_0 \frac{1}{z(1-z)} + g_1 \left(1+\frac{\log z}{z(1-z)}\right)
+ g_2 \left(1+\frac{\log(1-z)}{z(1-z)}\right)\;.
\eeq
We then want to construct the physical correlation functions which
must be single-valued on the whole Riemann sphere. It is therefore enough
to ensure this property at the two singular points 0 and 1.
It is easy to see that one can only have
\bea 
F_{AB}(z,\zb) &=& \af_0 F^{(0)}_A(z) F^{(0)}_B(\zb) + \af_1 (F^{(0)}_A(z) 
F^{(1)}_B(\zb) +  F^{(1)}_A(z) F^{(0)}_B(\zb))\nonumber \\
&+& \af_2 (F^{(0)}_A(z)
F^{(2)}_B(\zb) +  F^{(2)}_A(z) F^{(0)}_B(\zb))\;,
\eea
where the constants $\af_i$ are independent of the indices $A, B$, and
\bea 
&F^{(0)}_1(z)=\frac{z-2}{4\ep\ga (1-z)}\;,\;
F^{(0)}_2(z)=\frac{2 z-1}{4\ep\ga z(1-z)}\;,\;
F^{(0)}_3(z)=\frac{1}{z}&\nonumber\\
&F^{(1)}_1(z)=\frac{1}{4\ep\ga} \left(-1+\frac{z-2}{1-z}\log z\right)\;,\;
F^{(1)}_2(z)=\frac{1}{4\ep\ga z(1-z)} \left(1-z+(2 z-1) \log z\right)\;,&
\nonumber\\
&F^{(1)}_3(z)=\frac{1}{2 z} \left( z-1 +2\log z\right)\;,\;
F^{(2)}_1(z)=\frac{1}{4\ep\ga} \left( \frac{z}{z-1} +\frac{2-z}{1-z} \log(1-z)
\right)\;,&\nonumber\\
&F^{(2)}_2(z)=\frac{1}{4\ep\ga z(1-z)}\left(z+(1-2 z)\log(1-z)\right)\;,\;
F^{(2)}_3(z)=\frac{1}{2 z}\left( \frac{z(z-2)}{1-z}-2\log(1-z)\right).&
\eea

This form of $F_{AB}$ differs from the usual diagonal one
one finds for minimal models. Here there are three so far 
unconstrained constants (one constant is an overall 
normalization), even  after requiring monodromy invariance around the
three singular points.

We still have to require crossing symmetry among the correlators:
\beq
F^{i j k l}(z,\zb)=F^{i k j l}(1-z,1-\zb) \;,\;\; 
F^{i j k l}(z,\zb)=z^{-1/x} \zb^{-1/x} F^{i l k j}(1/z,1/\zb)\;.
\eeq
These conditions imply that $\af_1=\af_2=0$. These correlators are thus free
from logarithms. 

At level one the free fields $(\ga,\psi_-,\psi_+,\eta)$ provide a 
representation $[0,\frac{1}{2}]$ of the affine algebra $osp(2/2)_1$.
The corresponding correlation functions are then easily
calculated by making use of Wick's theorem. They agree with the results
obtained from the above calculation, and are free of logarithms. 

We then looked at $x=1/2$ ($k=3/2$) correlators for which we 
explicitly found the functions $F^{(i)}_A$. Again the physical correlators
were free from logarithms. It is possible to show that such is the case
for all  non-vanishing integer values of  $x^{-1}$. The solutions
$F^{(i)}_A$ have the same form as those for $x=1$ and $x=1/2$, 
namely a combination 
of rational functions of $z$ and of $\log z$ and $\log (1-z)$.
We were not able to  determine
the connection matrices between  functions  with arguments 
$z$, $1-z$ and $1/z$. The limiting values for those 
obtained in \cite{dotfat} depend on the way the limit is taken.
But it is enough to know the form 
of these matrices to implement the  crossing symmetries. In doing so
we also used the matrices connecting the tensors $I_a$ with transposed indices,
and which can be found in the appendix.  We found that $\af_1=\af_2=0$.
Note that the values of $k$ considered here are exactly those of equation
(\ref{restri}). We comment on this result in the conclusion.

From the four-point correlation functions we can obtain information 
about the fusion rules for the $[0,1/2]$ representation. As the contribution 
from the conformal blocks 1 and 2 vanishes we infer that the corresponding
operators do not appear in the OPE $\phi_{[0,1/2]}(z) \phi_{[0,1/2]}(w)$.
The only primary field appearing in their OPE corresponds to the operator
$\phi_t$, which is now the identity. 

\subsection{Logarithms for generic $x$}

We now consider the generic case where $ 2 x^{-1}$ is not an integer.
Contrary to what happens in the preceding section, it does not
seem possible to isolate a solution which is regular at both $z=0$ and $z=1$.
A solution regular at 0 will have a logarithmic singularity at $z=1$
and vice-versa. This means that there is no way to get rid of logarithms
if the correlators are not vanishing. 
We expect that  there should be logarithmic operators, and the knowledge
of the four-point functions allows one to find the OPE's of the fields 
we consider, including the logarithmic ones. 

\subsection{Squared logarithms}

We now turn to the third and last possible scenario 
where $2  x^{-1}$ is an integer
but $x^{-1}$ is not an integer: $x^{-1}=n+1/2$, $n\in \bZ$. 
The theory of fuchsian differential equations indicates that there
might be a solution with a $\log^2 z$ dependence. A careful analysis
shows that this is indeed the case. And again it seems impossible
to isolate a logarithm-free solution. However one has to modify the 
form of $F_{AB}$. Write the solutions as
\bea
&F_A^{(0)}(z)=F_A(z)\;,\;\;
F_A^{(1)}(z)=F_A(z) \log z + H_A(z)\;,&\nonumber \\
&F_A^{(2)}(z)=F_A(z) \log^2 z + 2 H_A(z) \log z  + G_A(z)\;.&
\eea
Then the monodromy invariant combinations around $z=0$
are 
\bea
F_{AB}(z,\zb)&=& \af_0 F_A^{(0)}(z) F_B^{(0)}(\zb) + \af_1 \left(F_A^{(0)}(z)
F_B^{(1)}(\zb) + F_A^{(1)}(z) F_B^{(0)}(\zb)\right) \nonumber\\
&+& \af_2 \left(F_A^{(0)}(z) F_B^{(2)}(\zb) + F_A^{(2)}(z) F_B^{(0)}(\zb)
+ 2 F_A^{(1)}(z) F_B^{(1)}(\zb) \right)\; .
\eea
We expect such a form to be also invariant around 1 without any additional 
constraint on the $\af_i$. Again, because there is no solution free of 
logarithms one expects a $\log^2$ dependence if the physical correlators 
do not vanish.

\section{Conclusion}

The current superalgebra  approach to gaussian disordered systems provides
a unifying framework for determining exponents of the models and
correlation functions. The logarithmic dependence of correlation functions
at critical points
is traced back to indecomposable representations which are common in
superalgebras. {\it Chiral} correlation functions possess a logarithmic
dependence. However for values of the level $k$ for which the theories
have good locality properties, the physical operators do not have a logarithmic
dependence. This conclusion was not apparent  from the start; it came out 
of imposing crossing symmetry. It is not clear at this stage what is the 
significance of the logarithmic operators in the context of the disordered 
systems. If they have vanishing dimension, as is the case here, they might be
related to the density of states.  

The correlation functions with $\log$ and
$\log^2$ dependence provide examples of 
correlators whose existence was predicted  in \cite{gurarie}.
The negative conformal weights are natural in this treatment of disorder
because one is considering non-unitary theories at $c=0$.
It should be also possible to obtain similar
results by considering $WZW$ models of cosets of Kac-Moody superalgebras.

\bigskip\ {\bf Acknowledgements:} It is a pleasure to thank D.~Bernard for
many suggestions. We also thank M. Bauer, P. Di Francesco for discussions.
Z.M. thanks R. Balian and J.M. Luck for discussions, and D.S. thanks
P. Sorba and L. Frappat. We are grateful to S. Nonnenmacher for  
many discussions. 

\section{Appendix}

The non-vanishing components of the three 
invariant tensors are equal to the components of the three vectors:
\bea
I_1&=& (1144)+(1234)4  \ep\ga+(1324)4  \ep\ga-(1414)+(2143)4  \ep\ga 
\nonumber\\&+&(2233)16 \ep^2\ga^2+
(2323)16 \ep^2\ga^2-(2413)4  \ep\ga+(3142)4  \ep\ga 
+(3232)16 \ep^2\ga^2 \nonumber \\
&+& (3322)16 \ep^2\ga^2 -(3412)4  \ep\ga-(4141)
-(4231)4  \ep\ga-(4321)4  \ep\ga+(4411)\\
I_2&=& (1234)4  \ep\ga-(1243)4  \ep\ga+(1324)4  \ep\ga-(1342)4  
\ep\ga \nonumber\\
&-& (1414)+(1441)-(2134)4  \ep\ga +(2143)4  \ep\ga+(2233)32 \ep^2\ga^2
+(2323)16 \ep^2\ga^2 \nonumber\\
&+& (2332)16 \ep^2\ga^2-(2413)4  \ep\ga
+(2431)4  \ep\ga-(3124)4  \ep\ga+(3142)4  \ep\ga \nonumber\\
&+& (3223)16 \ep^2\ga^2+(3232)16 \ep^2\ga^2
+(3322)32 \ep^2\ga^2-(3412)4  \ep\ga + (3421)4  \ep\ga \nonumber\\
&+&(4114) - (4141)+(4213)4  \ep\ga
-(4231)4  \ep\ga+(4312)4  \ep\ga-(4321)4  \ep\ga\\
I_3&=&(1234)-(1243)+(1324)-(1342)+(1423)+(1432)-(2134)+(2143)
\nonumber\\ 
&+&(2233)8  \ep\ga +(2314)+(2323)8  \ep\ga+(2332)8  \ep\ga
-(2341)-(2413)+(2431) \nonumber \\
&-& (3124)+(3142)+(3214)+(3223)8  \ep\ga+(3232)8  \ep\ga-(3241)
+(3322)8  \ep\ga \nonumber \\
&-& (3412)+(3421)-(4123)-(4132)+(4213)-(4231)+(4312)-(4321)
\eea

The matrices $P$ and $Q$ are given by:
\beq
P=\left( \begin{array}{ccc}
1 & 0 & 0 \\
-2 & -3 & -\frac{1}{2\ep\ga} \\
4\ep\ga & 8\ep\ga & 1    \end{array} \right) \quad ,\quad
Q=\left( \begin{array}{ccc}
-1 & 0 & -\frac{1}{2\ep\ga} \\
2 & 1 & \frac{1}{2\ep\ga} \\
-4\ep\ga & -4\ep\ga & -1 \end{array} \right) \;.
\eeq

The following tensors are needed in order to implement the 
crossing-symmetries:
\beq
J^{\af_1\af_2\af_3\af_4}_A = (-1)^{\vep_{\alpha_2} \vep_{\af_3}} 
I^{\af_1\af_3\af_2\af_4}_A\;,\;\;
K^{\af_1\af_2\af_3\af_4}_A = (-1)^{\vep_{\af_2} ( \vep_{\af_3} +\vep_{\af_4})
+\vep_{\af_3}\vep_{\af_4}} I^{\af_1\af_4\af_3\af_2}_A\; .
\eeq
The $\vep_{\af_i}$ refers to the parity of the index $\af_i$.
One has the following decomposition:
\bea
& J_1 = - I_1\;,\;\; J_2 = -I_1+I_2 -4\ep\ga I_3\;,\;\; J_3 = -I_3 & \\
& K_1 = I_2 -4\ep\ga I_3 \;,\;\; K_2 = I_1 -4\ep\ga I_3 \;,\;\; 
K_3 = - I_3 & \; . 
\eea
We rewrite this compactly as $J=C_1\, I$ and $K=C_2\, I$ where $C_{1,2}$ are 
two matrices which are used to implement the crossing symmetry constraints.

\end{document}